\documentclass[a4paper]{jpconf}
\usepackage{graphicx}
\begin{document}
\title{Impact parameter dependence of collective flow and its disappearance for different mass asymmetries.}

\author{Supriya Goyal}

\address{Department of Physics, Panjab University, Chandigarh 160 014, India}

\ead{ashuphysics@gmail.com}

\begin{abstract}
We study the role of impact parameter on the collective flow and
its disappearance for different mass asymmetric reactions. The
mass asymmetry is varied from 0 to 0.7 keeping the total mass of
the system fixed. Our results clearly indicate a significant role
of impact parameter on the collective flow and its disappearance
for the mass asymmetric reactions. The impact parameter dependence
is also found to vary with mass asymmetry of the reaction.
\end{abstract}

\section{Introduction}
On the collision of two nuclei, the density and pressure increases
in the interaction region and at finite colliding geometries,
there is an inherent asymmetry in the pressure which results in a
collective transverse flow of matter toward the direction of
lowest pressure. This collective transverse flow is one of the
most extensively used observable to study the equation of state
(EoS) as well as in-medium nucleon-nucleon (nn) cross-section of
the nuclear matter. At the particular value of energy where the
attractive scattering (dominant at energies around 10 MeV/nucleon)
balances the repulsive interactions (dominant at energies around
400 MeV/nucleon), collective transverse flow in the reaction plane
disappears. This energy is termed as Balance Energy ($E_{bal}$)
\cite{1,2,3,4,5,6,7,8,9,10}.
\par
Both collective flow and $E_{bal}$ are found to be highly
sensitive towards the nn cross-section \cite{2,3,4,5,6,7,8,9,10},
size of the system ($A_{TOT} = A_{T}+A_{P}$; where $A_{T}$ and
$A_{P}$ are the masses of the target and projectile, respectively)
\cite{4}, mass asymmetry of the reaction ($\eta$ =
$\frac{A_{T}-A_{P}}{A_{T}+A_{P}}$) \cite{11}, nuclear matter
equation of state \cite{2,3,4,5,6,7,8,9,10}, incident energy (E
MeV/nucleon) \cite{12} as well as the colliding geometry
($\hat{b}$ = $\frac{b}{b_{max}}$; where $b_{max}$ = $R_{1} +
R_{2}$; $R_{i}$ is the radius of projectile or target)
\cite{2,5,7,9,10}. Several experimental and theoretical attempts
have been employed in order to explain and understand these
observations \cite{1,2,3,4,5,6,7,8,9,10,11,12}. Due to the
decrease in the compression reached in heavy-ion collisions with
increase in the impact parameter, $E_{bal}$ is found to increase
approximately linearly as a function of the impact parameter
\cite{13,14}. But most of the studies in the literature mainly
focus on the symmetric and nearly symmetric reactions. Our present
aim, therefore, is at least twofold. (1) To study the variation of
collective transverse flow with impact parameter for different
$\eta$ keeping the incident energy fixed. (2) To study the
sensitivity of collective transverse flow and its disappearance on
the mass asymmetry of the reaction at different impact parameters.
The Quantum Molecular Dynamics (QMD) model \cite{8,15,16} used for
the present analysis is explained in the section 2. Results and
discussion are presented in section 3 followed by summary in
section 4.

\section{\label{model}Description of the model}
In the QMD model, each nucleon propagates under the influence of
mutual two- and three-body interactions. The propagation is
governed by the classical equations of motion:
\begin{equation}
\frac {d\vec{r}_{i}}{dt} = \frac {dH}{d\vec{p}_{i}},
\end{equation}
\begin{equation}
\frac {d\vec{p}_{i}}{dt} = -\frac {dH}{d\vec{r}_{i}},
\end{equation}
where the Hamiltonian is given by
\begin{equation}
H=\sum_{i} \frac {\vec{p}_{i}^{2}}{2m_{i}} + V ^{tot}.
\end{equation}
 Our total interaction potential $V^{tot}$ reads as \cite{8,15,16}
\begin{equation}
V^{tot} = V^{Loc} + V^{Yuk} + V^{Coul},
\end{equation}
with
\begin{equation}
V^{Loc} = t_{1}\delta(\vec{r}_{i}-\vec{r}_{j})+
t_{2}\delta(\vec{r}_{i}-\vec{r}_{j})
\delta(\vec{r}_{i}-\vec{r}_{k}),
\end{equation}
\begin{equation}
V^{Yuk}=t_{3}e^{-|\vec{r}_{i}-\vec{r}_{j}|/m}/\left(|\vec{r}_{i}-\vec{r}_{j}|/m\right),
\end{equation}
with ${\it m}$ = 1.5 fm and $\it{t_{3}}$ = -6.66 MeV.
\par
The static (local) Skyrme interaction can further be parametrized
as:
\begin{equation}
U^{Loc}=\alpha\left(\frac{\rho}{\rho}_o\right)+
\beta\left(\frac{\rho}{\rho}_o\right)^{\gamma}.
\end{equation}
Here $\alpha, \beta$ and $\gamma$ are the parameters that define
equation of state. The parameters $\alpha$, $\beta$, and $\gamma$
in above Eq. (7) must be adjusted so as to reproduce the ground
state properties of the nuclear matter. The set of parameters
corresponding to different equations of state can be found in Ref.
\cite{16}. It is worth mentioning that as shown by Puri and
coworkers, Skyrme forces are very successful in the analysis of
low energy phenomena such as fusion, fission and
cluster-radioactivity, where nuclear potential plays an important
role \cite{sky}.
\begin{figure}[!t]
\centering \vskip - 1.0 cm
\includegraphics* [scale=0.7] {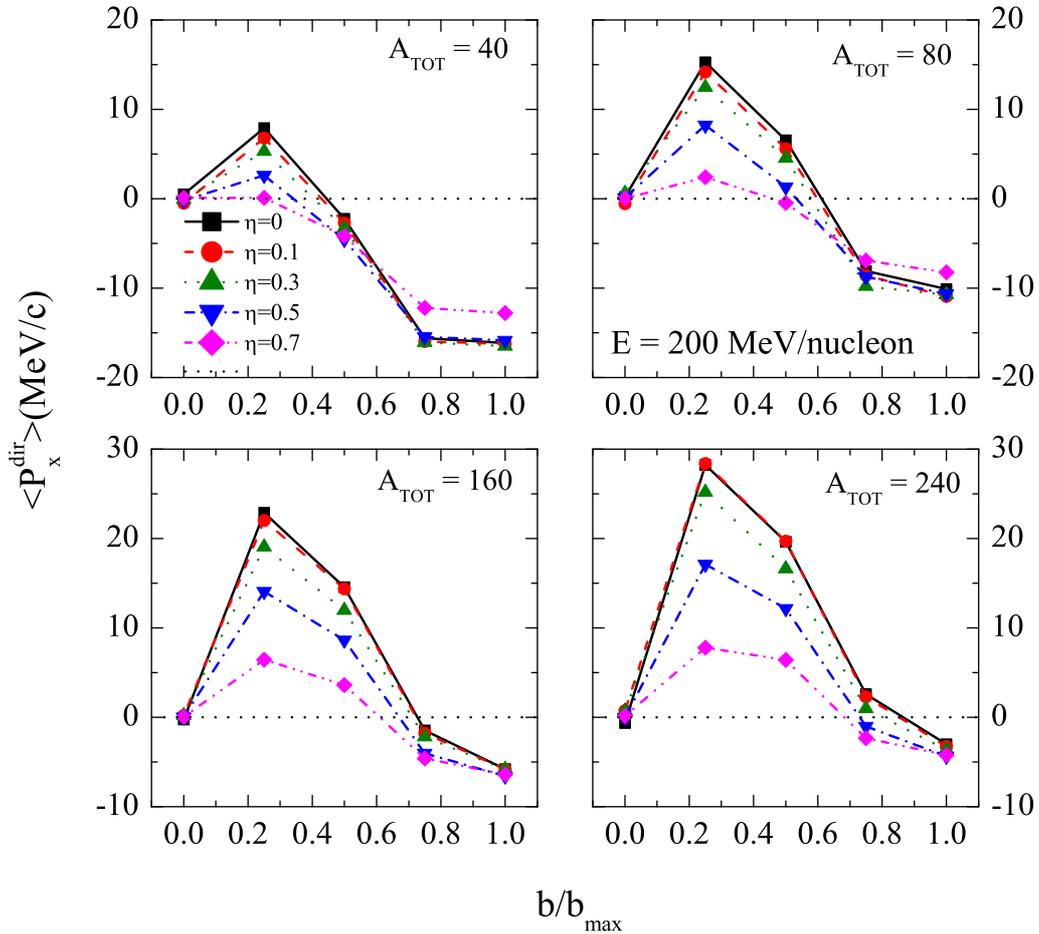}
\vskip -5.0 cm \caption {(Color online) The $<P^{dir}_{x}>$
(MeV/c) as a function of reduced impact parameter ($b/b_{max}$)
for different system masses. The results for different mass
asymmetries $\eta$ = 0, 0.1, 0.3, 0.5, and 0.7 are represented,
respectively, by the solid squares, circles, triangles, inverted
triangles, and diamonds. Results are at an incident energy of 200
MeV/nucleon.}
\end{figure}
\begin{figure}
\centering \vskip -7.5 cm
\includegraphics* [scale=0.7] {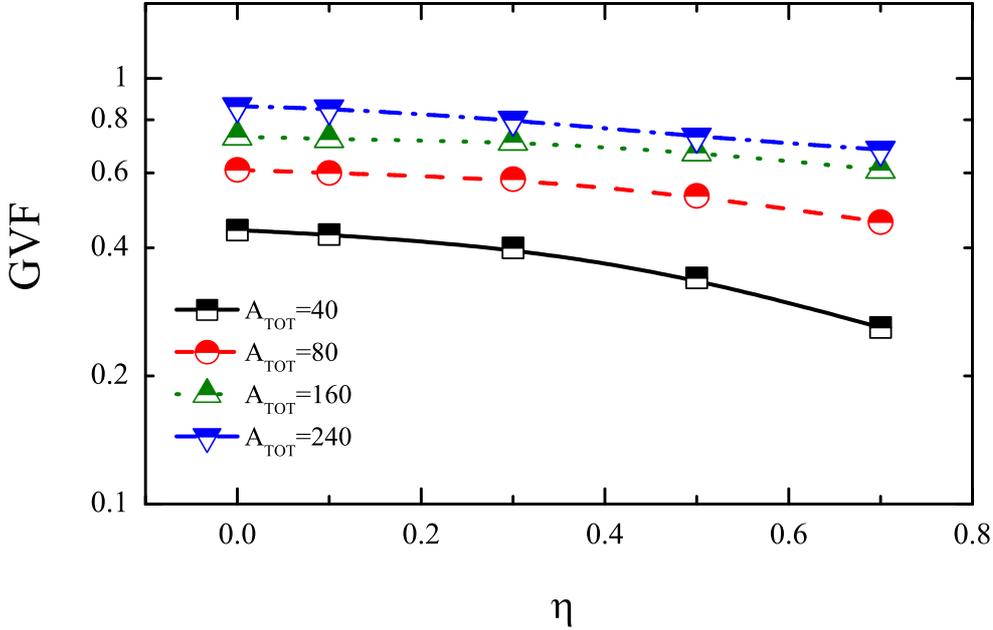}
\vskip -2.0 cm \caption {(Color online) The geometry of vanishing
flow (GVF) as a function of ${\eta}$ for different system masses.
The results for different system masses ($A_{TOT}$) = 40, 80, 160,
and 240 are represented, respectively, by the half filled squares,
circles, triangles, and inverted triangles.}
\end{figure}
\begin{figure}[!t]
\centering
\includegraphics* [scale=0.7] {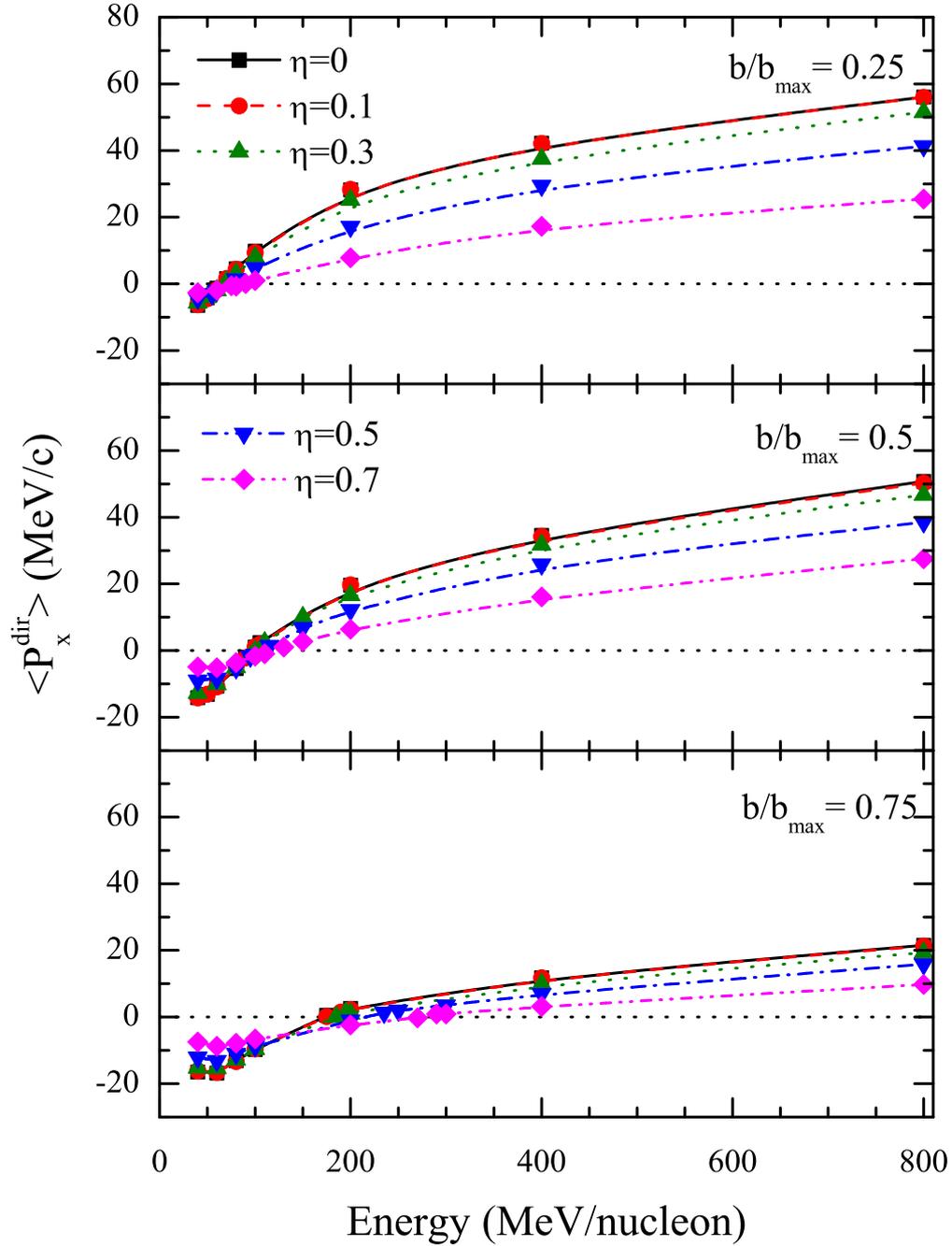}
\vskip -1.0 cm \caption {(Color online) The $<P^{dir}_{x}>$
(MeV/c) as a function of incident energy for system mass $A_{TOT}$
= 240. The results are shown for different mass asymmetries
($\eta$ = 0-0.7) at reduced impact parameters $b/b_{max}$ = 0.25
(upper panel), 0.5 (middle panel), and 0.75 (bottom panel). The
lines are only to guide the eye. Symbols have the same meaning as
in Fig. 1.}\label{fig3}
\end{figure}

\section{\label{results} Results and Discussion}
For the present study, we simulated the reactions of
$^{20}_{10}Ne+^{20}_{10}Ne$ ($\eta = 0$),
$^{17}_{8}O+^{23}_{11}Na$ ($\eta = 0.1$),
$^{14}_{7}N+^{26}_{12}Mg$ ($\eta = 0.3$),
$^{10}_{5}B+^{30}_{14}Si$ ($\eta = 0.5$), and
$^{6}_{3}Li+^{34}_{16}S$ ($\eta = 0.7$) for total mass ($A_{TOT}$)
= 40, $^{40}_{20}Ca+^{40}_{20}Ca$ ($\eta = 0$),
$^{36}_{18}Ar+^{44}_{20}Ca$ ($\eta = 0.1$),
$^{28}_{14}Si+^{52}_{24}Cr$ ($\eta = 0.3$),
$^{20}_{10}Ne+^{60}_{28}Ni$ ($\eta = 0.5$), and
$^{10}_{5}B+^{70}_{32}Ge$ ($\eta = 0.7$) for total mass
($A_{TOT}$) = 80, $^{80}_{36}Kr+^{80}_{36}Kr$ ($\eta = 0$),
$^{70}_{32}Ge+^{90}_{40}Zr$ ($\eta = 0.1$),
$^{54}_{26}Fe+^{106}_{48}Cd$ ($\eta = 0.3$),
$^{40}_{20}Ca+^{120}_{52}Te$ ($\eta = 0.5$), and
$^{24}_{12}Mg+^{136}_{58}Ce$ ($\eta = 0.7$) for total mass
($A_{TOT}$) = 160, and $^{120}_{52}Te+^{120}_{52}Te$ ($\eta = 0$),
$^{108}_{48}Cd+^{132}_{56}Ba$ ($\eta = 0.1$),
$^{84}_{38}Sr+^{156}_{66}Dy$ ($\eta = 0.3$),
$^{60}_{28}Ni+^{180}_{74}W$ ($\eta = 0.5$), and
$^{36}_{18}Ar+^{204}_{82}Pb$ ($\eta = 0.7$) for total mass
($A_{TOT}$) = 240. The impact parameter is varied from b/b$_{max}$
= 0 to 1 in small steps of 0.25. The charges are chosen in a way
so that colliding nuclei are stable nuclides. A soft equation of
state with isotropic energy dependent cugnon cross-section
(labeled as Soft$^{iso}$) is used for the present calculations.
\par
The balance energy ($E_{bal}$) is calculated using the {\it
directed transverse momentum $<P^{dir}_{x}>$}, which is defined
as:
\begin{equation}
\langle P_{x}^{dir}\rangle=\frac{1}{A}\sum_i {\rm
sign}\{Y(i)\}~{\bf{p}}_{x}(i),
\end{equation}
where $Y(i)$ and ${\bf{p}}_{x}(i)$ are the rapidity distribution
and transverse momentum of $i^{th}$ particle, respectively.
\par
In Fig. 1, we display at a fixed energy, the $<P^{dir}_{x}>$ as a
function of reduced impact parameter ($b/b_{max}$) for ${\eta}$ =
0-0.7, keeping total system mass fixed as 40, 80, 160, and 240.
All reactions are followed till 200 fm/c, where $<P^{dir}_{x}>$
saturates. In all cases $<P^{dir}_{x}>$ first increases with
increase in impact parameter, reaches a maximal value and after
passing through a zero at some intermediate value of impact
parameter, attains negative values. The trend is uniform
throughout the mass asymmetry range. The value of impact parameter
at which $<P^{dir}_{x}>$ attains a zero (which is termed as
Geometry of Vanishing Flow (GVF) \cite{17}) varies with ${\eta}$
and $A_{TOT}$. For lighter systems and larger ${\eta}$, the value
of GVF is smaller compared to the heavier systems and smaller
${\eta}$.
\begin{figure}[!t]
\centering
\includegraphics* [scale=0.7]{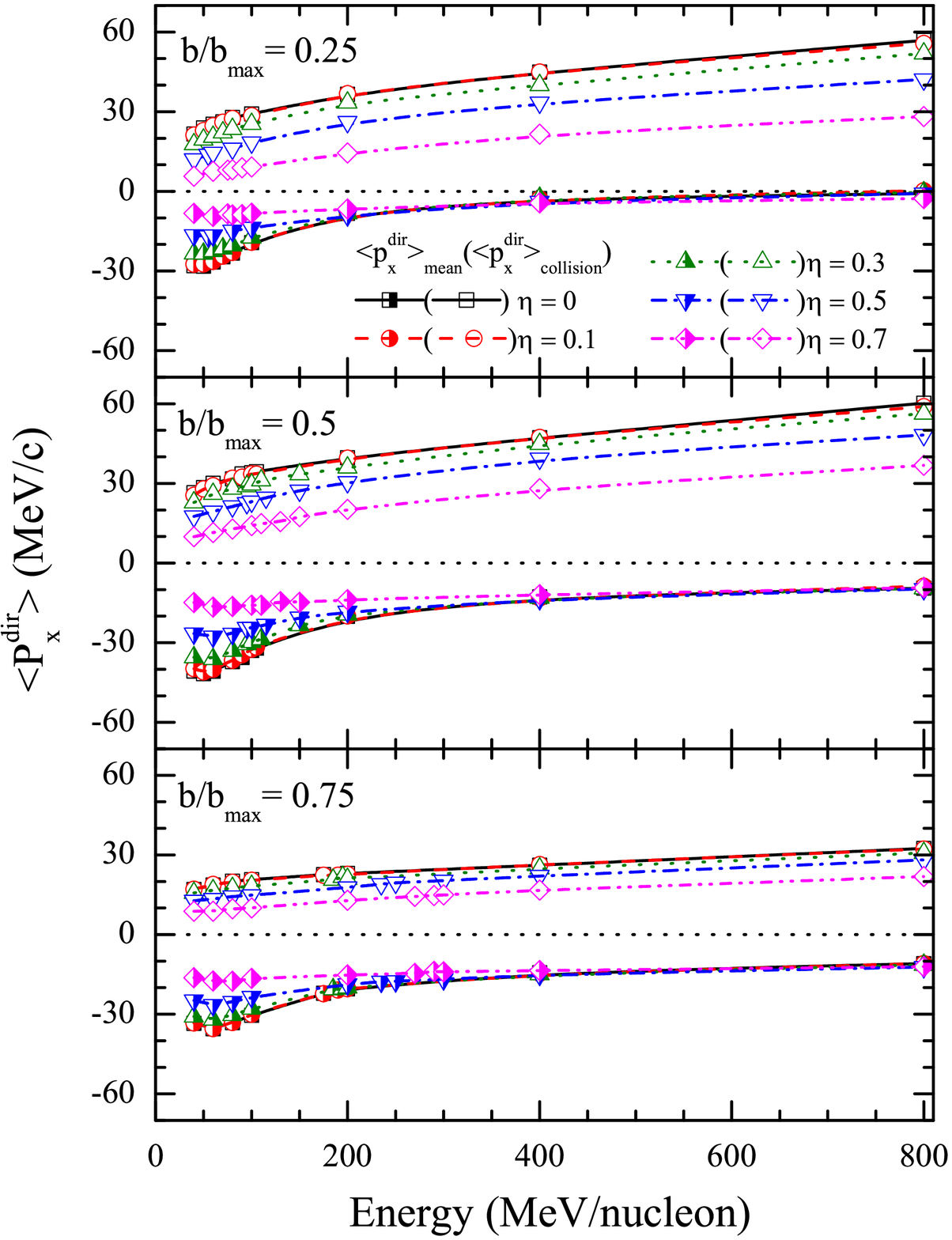}
\vskip -0.5 cm \caption {(Color online) The decomposition of
$<P^{dir}_{x}>$ displayed in Fig. 3 into mean field (half filled
symbols) and collision part (open symbols) as a function of
incident energy.}\label{fig4}
\end{figure}
\begin{figure}
\centering  \vskip -7.5 cm
\includegraphics* [scale=0.7]{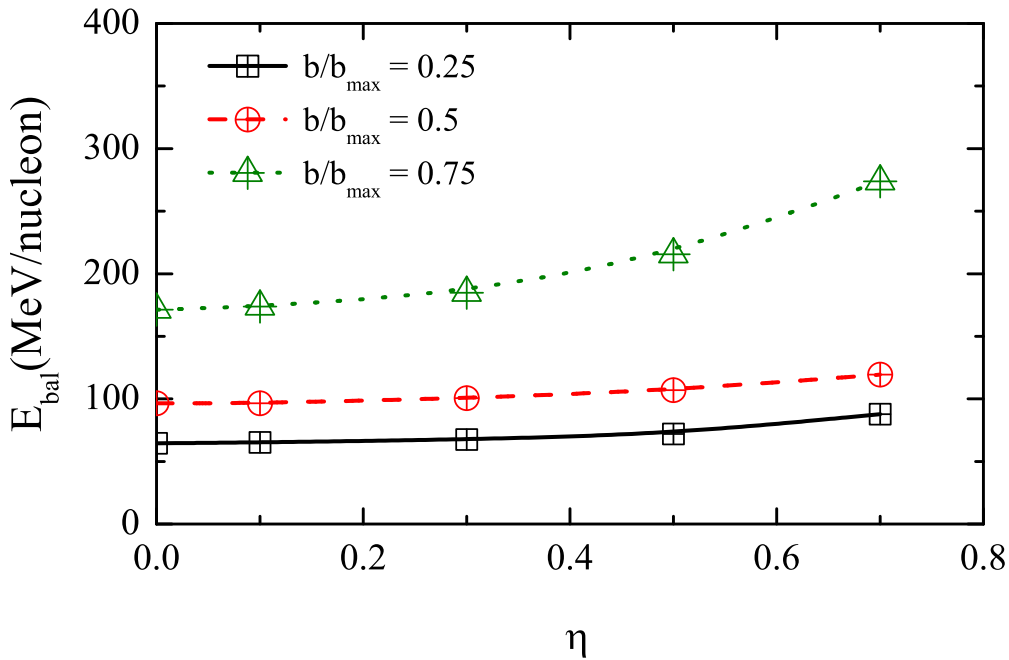}
\vskip -2.0 cm \caption {(Color online) The $E_{bal}$ as a
function of $\eta$ for system mass $A_{TOT}$ = 240. The results
for different impact parameters  $b/b_{max}$ = 0.25, 0.5, and 0.75
are represented, respectively, by the crossed squares, circles,
and triangles. Lines are to guide the eye.}\label{fig5}
\end{figure}
\begin{figure}
\centering \vskip -1.0 cm
\includegraphics* [scale=0.7] {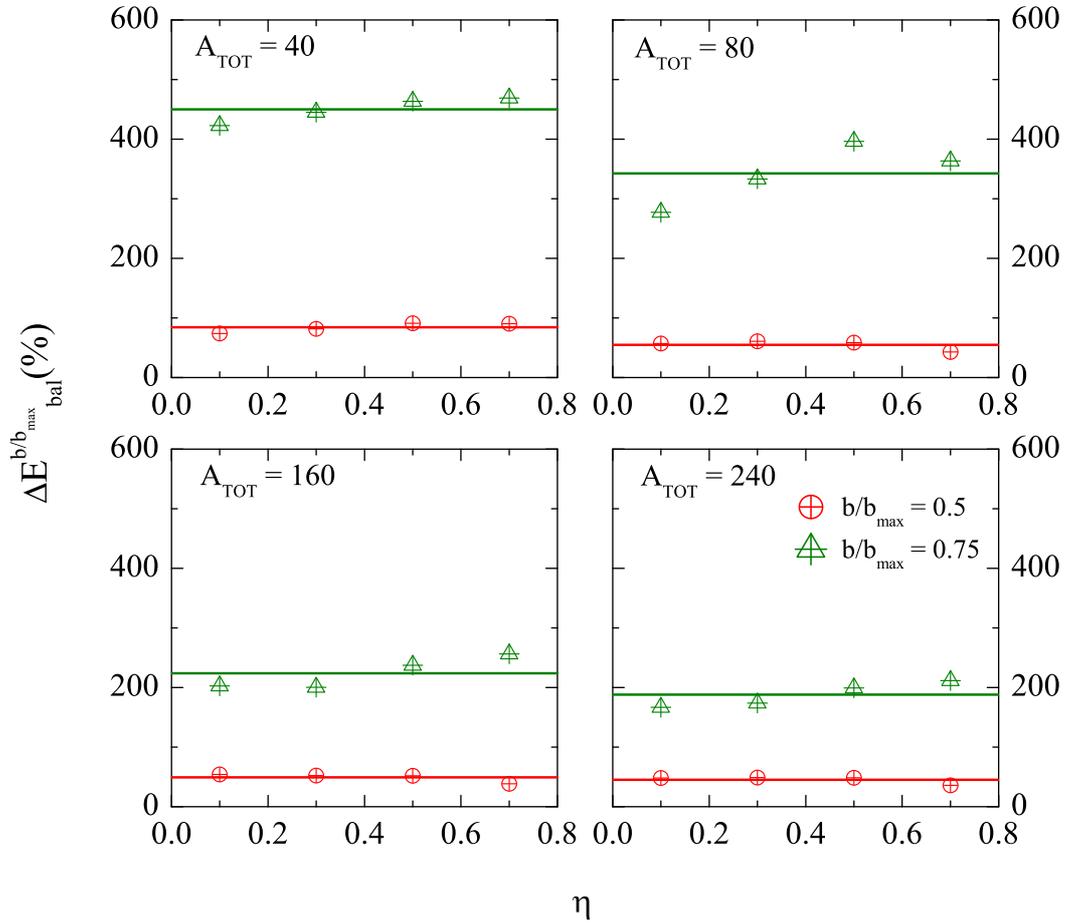}
\vskip -5.0 cm \caption {(Color online) The percentage difference
$\Delta E_{bal}^{b/b_{max}}$(\%) as a function of ${\eta}$ for
different system masses. The results of the percentage difference
for different colliding geometries $b/b_{max}$ = 0.5 and 0.75 are
represented, respectively, by the crossed circles and triangles.
Horizontal lines represent the mean value of $\Delta
E_{bal}^{b/b_{max}}$(\%) for each $b/b_{max}$.}\label{fig6}
\end{figure}
\begin{figure}
\centering \vskip - 1.0 cm
\includegraphics* [scale=0.7] {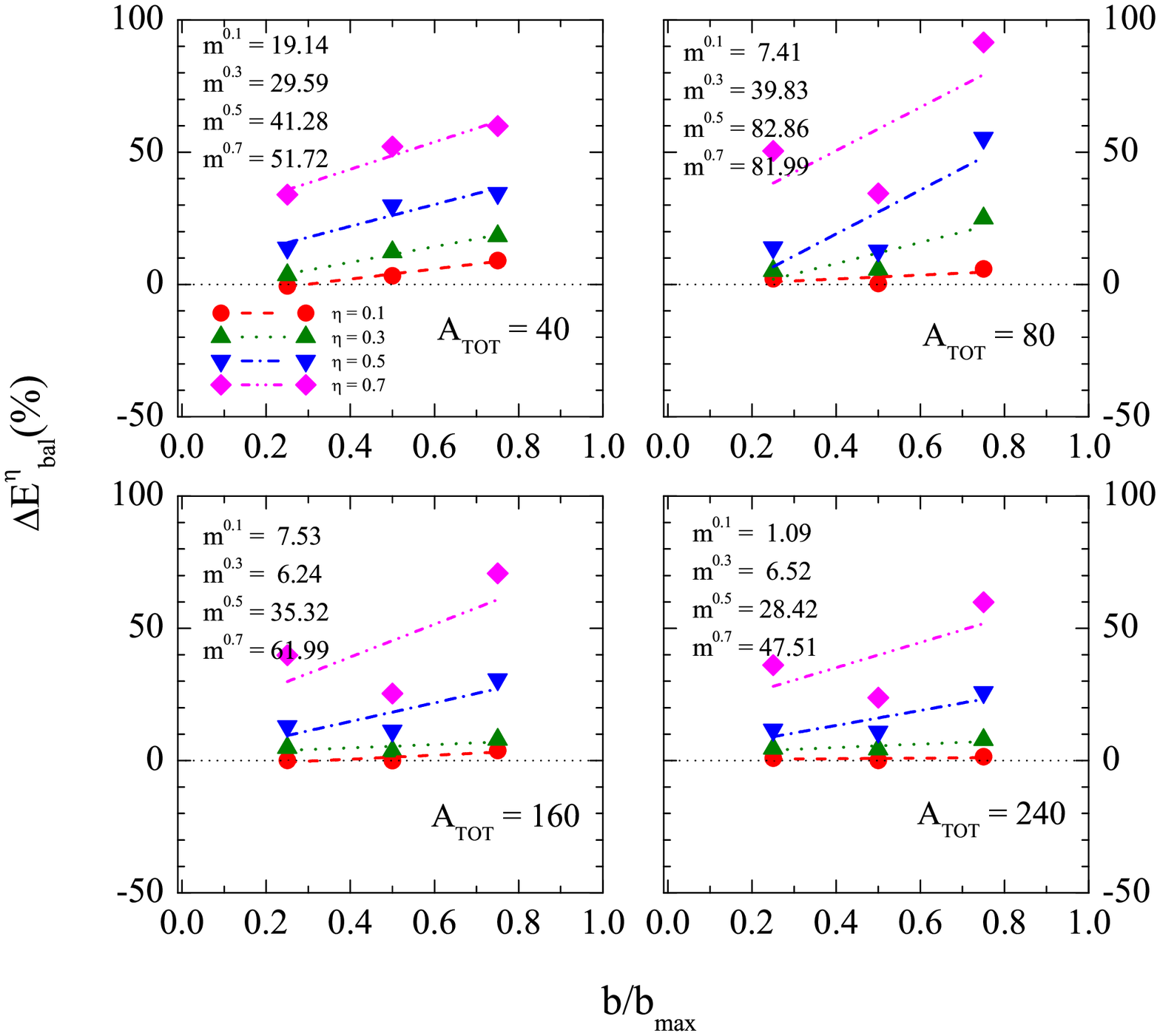}
\vskip -5.0 cm \caption {(Color online) The percentage difference
$\Delta E_{bal}^{\eta}$(\%) as a function of $b/b_{max}$ for
different system masses. The results of the percentage difference
for different asymmetries $\eta$ = 0.1, 0.3, 0.5, and 0.7 are
represented, respectively, by the solid circles, triangles,
inverted triangles, and diamonds. Lines are the linear fits
($\propto m\frac {b}{b_{max}}$); {\it m} values without errors are
displayed.}\label{fig7}
\end{figure}
\par
In Fig. 2, the variation of GVF as a function of ${\eta}$ is
displayed for different system masses. The percentage variation in
GVF while going from ${\eta}$ = 0 to 0.7 is  -40.9\%, -24.59\%,
-16.44\%, and -20.93\%, respectively for $A_{TOT}$ = 40, 80, 160,
and 240. It is clear from the figure that the effect of mass
asymmetry of the reaction on GVF decreases with increase in system
mass. This is similar to as predicted for $E_{bal}$ \cite{11}. Due
to the decrease in nn collisions with increase in ${\eta}$ and
impact parameter and increase in Coulomb repulsion with increase
in $A_{TOT}$, the $E_{bal}$ increases with increase in ${\eta}$
and impact parameter, while it decreases with increase in
$A_{TOT}$. Since the present study is at fixed incident energy,
therefore, the value of impact parameter, where flow vanishes,
decreases as ${\eta}$ increases.
\par
In Fig. 3, we display $<P^{dir}_{x}>$ as a function of incident
energy ranging between 40 MeV/nucleon and 800 MeV/nucleon, for
different mass asymmetric reactions keeping the total mass of the
system fixed as 240. The results are shown for different impact
parameters. In all the cases, transverse momentum is negative at
lower incident energies which turns positive at relatively higher
incident energies. The value of the abscissa at zero value of
$<P^{dir}_{x}>$ corresponds to the balance energy. The figure
indicates that (i) for all values of $\eta$ and impact parameters,
the transverse momentum increases monotonically with increase in
the incident energy. The increase is sharp at smaller incident
energies compared to higher incident energies where it starts
saturating. (ii) due to decrease in overlap volume and hence
number of collisions with increase in $\eta$, the transverse
momentum starts suppressing as $\eta$ increases and hence the
balance energy increases with increase in $\eta$. (iii) the
variation in $<P^{dir}_{x}>$ with $\eta$ decreases with increase
in impact parameter. But with increase in impact parameter the
balance energy increases for all values of $\eta$.
\par
The above mentioned findings can be understood by decomposing the
total transverse momentum into contributions due to mean field and
two- body nn collisions as shown in Fig. 4. The symbols are
explained in the caption of the figure. One notices that mean
field flow increases up to couple of hundred MeV/nucleon and then
saturates. In the lower incident energy region, the flow due to
mean field is smaller for larger asymmetries compared to smaller
asymmetries. At higher incident energies the mean field flow is
nearly independent of $\eta$. Similar behavior is seen for other
impact parameters also. The flow due to binary nn collisions,
however, increases with increase in incident energy for all
$\eta$, but as the impact parameter increases, the nn collision
flow starts saturating. Obviously as explained earlier, the flow
due to binary nn collisions decreases with increase in $\eta$.
\par
In Fig. 5, we display $E_{bal}$ as a function of $\eta$ for
$b/b_{max}$ = 0.25, 0.5, and 0.75, keeping $A_{TOT}$ fixed as 240.
Various symbols are explained in the caption of the figure. From
the figure, we see that $E_{bal}$ increases with increase in
$\eta$ for all values of $b/b_{max}$. Also the $\eta$ dependence
of $E_{bal}$ increases with increase in impact parameter.
\par
In Fig. 6, we display the percentage change in balance energy
$\Delta E_{bal}^{b/b_{max}}$(\%), defined as $\Delta
E_{bal}^{b/b_{max}}$(\%) =
(($E_{bal}^{b/b_{max}\neq0.25}$-$E_{bal}^{b/b_{max}=0.25}$)/$E_{bal}^{b/b_{max}=0.25}$)$\times$100
as a function of $\eta$. Lines represents the mean value of
variation. It is clear from the figure that the effect of impact
parameter variation is almost uniform throughout the asymmetry
range for every fixed system mass. We also found that the mean
variation decreases with increase in $A_{TOT}$. Clearly, the
effect of impact parameter variation is independent of ${\eta}$.
\par
In Fig. 7, we display the percentage difference $\Delta
E_{bal}^{\eta}$(\%) defined as $\Delta E_{bal}^{\eta}$(\%) =
(($E_{bal}^{\eta\neq0}$ -
$E_{bal}^{\eta=0}$)/$E_{bal}^{\eta=0}$)$\times$100 as a function
of reduced impact parameter ($b/b_{max}$). Lines are the linear
fits ($\propto m\frac {b}{b_{max}}$). It is clear from the figure
that the effect of the asymmetry variation increases with increase
in the impact parameter for each mass range. This is due to the
fact that with increase in impact parameter, the number of binary
nn collisions decreases and the increase of mass asymmetry further
adds the same effect.

\section{\label{summary} Summary}
We presented a detailed study on the role of impact parameter on
the collective flow and its disappearance for different mass
asymmetric reactions using the quantum molecular dynamics model.
For the present study, the mass asymmetry of the reaction is
varied from 0 to 0.7 by keeping the total mass of the system
fixed. A significant role of impact parameter on the collective
flow and its disappearance for the mass asymmetric reactions is
seen. The impact parameter dependence is also found to vary with
mass asymmetry of the reaction.

\section{Acknowledgments}
Author is thankful to Council of Scientific and Industrial
Research (CSIR) for providing the Junior Research Fellowship.

\smallskip
\end{document}